\documentclass[prl,superscriptaddress,graphix,twocolumn,reprint]{revtex4}

\usepackage{graphicx}
\usepackage{dcolumn}
\usepackage{bm}

\begin{document}

\preprint{Preprint submitted to Nature Photonics}

\title{Multiphoton-excited DUV photolithography for 3D nanofabrication}



\author{Atsushi Taguchi}
\email[]{taguchi@es.hokudai.ac.jp}
\affiliation{Department of Applied Physics, Osaka University, Suita, Osaka 565-0871, Japan}
\altaffiliation{Present address: Research Institute for Electronic Science, Hokkaido University, Sapporo, Hokkaido 001-0020, Japan}
\author{Atsushi Nakayama}
\affiliation{Department of Applied Physics, Osaka University, Suita, Osaka 565-0871, Japan}
\affiliation{Advanced Photonics and Biosensing Open Innovation Laboratory, AIST-Osaka University, Suita, Osaka 565-0871, Japan}
\author{Ryosuke Oketani}
\affiliation{Department of Applied Physics, Osaka University, Suita, Osaka 565-0871, Japan}
\author{Satoshi Kawata}
\affiliation{Department of Applied Physics, Osaka University, Suita, Osaka 565-0871, Japan}
\affiliation{Serendip Research, Suita, Osaka 565-0871, Japan}
\author{Katsumasa Fujita}
\email[]{fujita@ap.eng.osaka-u.ac.jp}
\affiliation{Department of Applied Physics, Osaka University, Suita, Osaka 565-0871, Japan}
\affiliation{Advanced Photonics and Biosensing Open Innovation Laboratory, AIST-Osaka University, Suita, Osaka 565-0871, Japan}
\affiliation{Transdimensional Life Imaging Division, Institute for Open and Transdisciplinary Research Initiatives, Osaka University, Suita, Osaka 565-0871, Japan}




\begin{abstract}
Light-matter interactions in the deep ultraviolet (DUV) wavelength region exhibits a variety of optical effects such as luminescence, photoisomerization, and polymerization in many materials.
Despite the rich photochemistry and high spatial resolution due to the short wavelength, the notorious lack of DUV-compatible optical components and devices precludes use of DUV light in microscopy and lithography as a routine laboratory tool.
Here, we present the use of two-photon excitation with visible laser light to realizes photo-polymerization of molecules with an excitation energy equivalent to DUV light.
Using standard optics for visible light, methacrylate oligomers were polymerized with 400\,nm femtosecond pulses without any addition of photo-initiators and sensitizers.
By scanning the laser focus in 3D, a series of fine 3D structures were created with the smallest resolved line-space features of 80\,nm.
We found DUV polymerizations induced by two-photon absorption is surprisingly efficient and requires laser intensity only on the order of 100\,kW/cm$^2$.
With the variety of successful demonstrations including organic- and inorganic-material-made-structures presented, our direct nano-3D-printing method would be a valuable tool for nanofabrication in 3D.


\end{abstract}


\maketitle 


The spatial resolution in optical microscopy and lithography is determined by the wave nature of light, namely, the half the wavelength.
As a result of the successive shortening of the operating wavelength, the advanced deep UV (DUV) lithography machines  is operated with the 193\,nm wavelength of argon fluoride laser \cite{Wagner:2010kb,Totzeck:2007tk} and electron beam lithography (EBL) makes use of even shorter de Broglie wave of electrons to resolve nanoscale features.
The major limitation facing when using such short wavelengths is the lack of optically transparent materials and compatible devices in the short wavelength region below 340\,nm.



As a solution, we present the use of two-photon excitation with visible wavelength ($\lambda$400\,nm) to realize energetically equivalent DUV excitation ($\lambda$200\,nm) for photo-polymerization.
Previous studies in fluorescence microscopy demonstrate two- \cite{Yamanaka:2015dw} and three-photon \cite{Maiti:1997wa} excitation is capable to access high energy molecular transition at DUV using standard optics for visible.
Thanks to the strong nonlinearly involved in two-photon polymerization process \cite{Kawata:2001hw}, a superior spatial resolution is achievable beyond the diffraction limit of the light, which can be comparable to DUV lithography and EBL \cite{Gan:2013ed}.
The nonlinearity also offers 3D fabrication capability \cite{Maruo:1997ct,Kawata:2001hw}, the feature not accessible by DUV lithography technology based on one-photon excitation.

\begin{figure*}
\begin{center}
\includegraphics[]{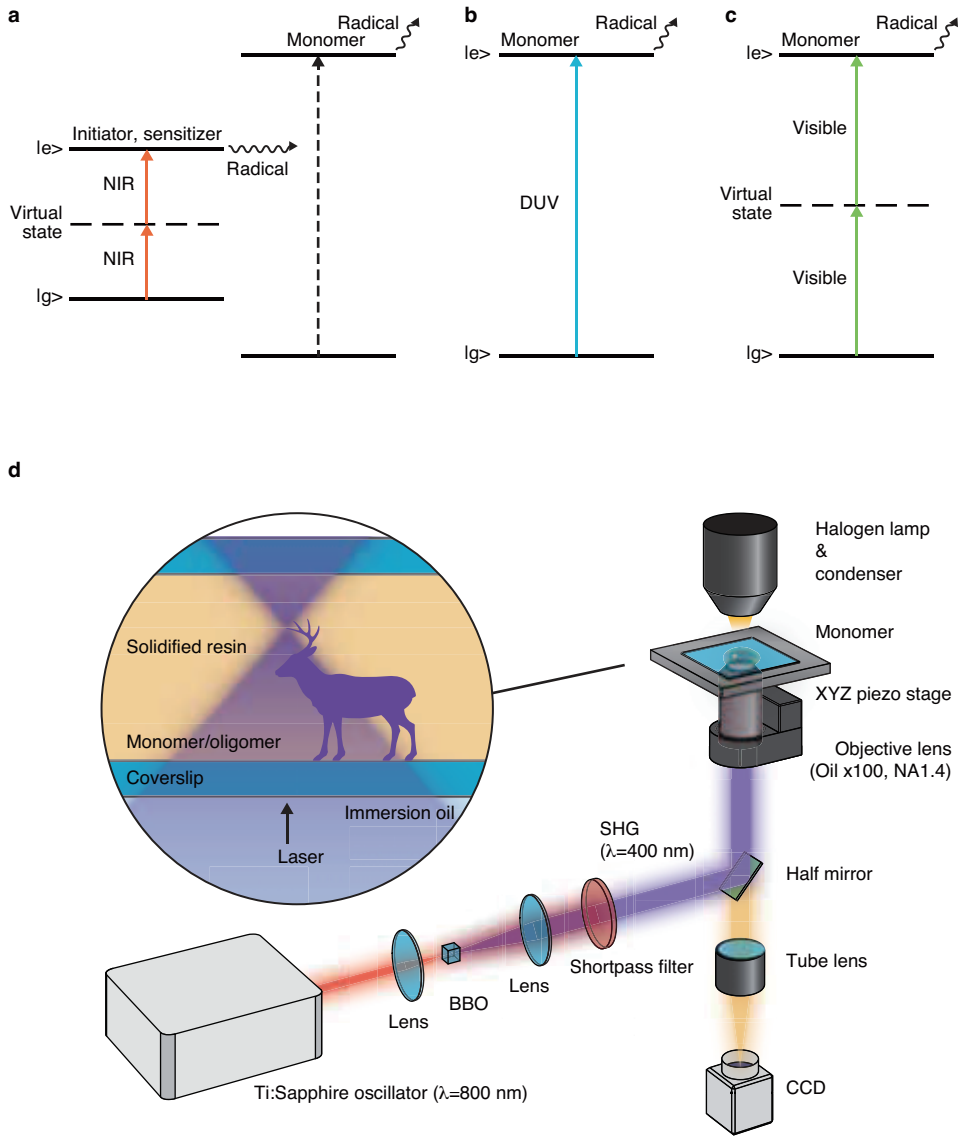}
\caption{\label{f1}
Two-photon polymerization in deep ultraviolet (DUV).
a--c, Energy diagrams of two-photon polymerization using NIR excitation (a), one-photon-absorption polymerization using DUV excitation (b), and two-photon polymerization using visible excitation (c).
d, Experimental setup (details in the Methods). Femto-second pulsed laser of Ti:sapphire oscillator (80\,fs, 82\,MHz) with a wavelength of 800\,nm is frequency-doubled using a $\beta$-BaB$_{2}$O$_{4}$ (BBO) crystal and subsequently focused into liquid monomer by a high NA objective.
}
\end{center}
\end{figure*}

Traditionally, TPP technique employs a Ti:sapphire-based femtosecond oscillator operating at NIR wavelength ($\sim\lambda$800\,nm) as an excitation laser.
The use of short excitation wavelength at 400\,nm in TPP, as presented here, changes the picture of photochemistry.
To explain this, we illustrated energy diagrams for photo-excitation and radical generations in Figs.~\ref{f1}(a)--(c).
As illustrated in Fig.~\ref{f1}(a), excitation energy given by two-photon absorption of NIR pulses is too low to directly excite monomers to the electronic excited state.
To trigger polymerization chain reactions, photo-initiators and sensitizers that have a transition energy corresponding to the two-photon energy of NIR light are added in monomer solution.
However, the addition of initiators, for example, in fabricating gelatin hydrogel scaffolds for tissue engineering shows problematic cytotoxicity during cell culturing \cite{Li:2013bd,Williams:2005ds}, suggesting the addition of initiators can affect properties of the fabricated devices.
On the other hand, the wavelength shorter than 220\,nm covers absorption bands of double bonds including C=C, C=O, benzene ring, and so on (Supplementary Table S1).
As shown in Fig.~\ref{f1}(b), DUV wavelength as short as 200\,nm allows direct excitation of the chemical bonds natively existing in organic molecules, which enables polymerization without using initiators.
Excited with energetically equivalent two-photon absorption (Fig.~\ref{f1}(c)), initiator-free DUV polymerization can be realized with visible excitation wavelength.
To the best of our knowledge, the present work is the first to use natively existing double bonds in organic molecules for TPP nanofabrications.

Figure \ref{f1}(d) illustrates the optical setup (details in the Methods).
Briefly, femtosecond pulse laser from Ti:sapphire oscillator operated at NIR wavelength ($\lambda$800\,nm) is frequency-doubled to produce visible femtosecond pulse ($\lambda$400\,nm) and focused by a high NA objective (NA$=$1.4) into a sample cell containing liquid monomer/oligomer.
The position of the sample cell is scanned three-dimensionally in $XYZ$ with respect to the position of the laser focus spot to pile up voxels to form a structure.

\begin{figure*}
\begin{center}
\includegraphics*[]{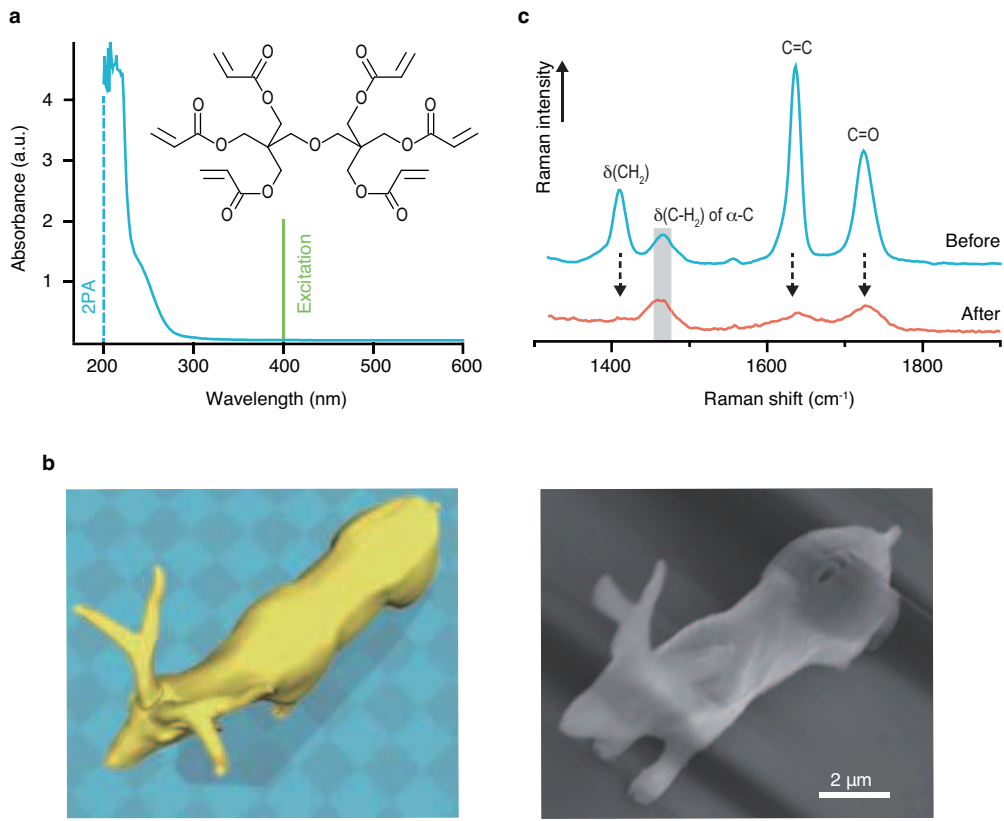}
\caption{\label{f2}
Two-photon polymerization of acrylate oligomer.
(a) Absorption spectrum of di-pentaerythritol polyacrylate (DPHA).
(b) Surface rendering of a deer made by computer-aided-design (CAD) software (left) and the scanning electron micrograph of the fabricated microsculpture (right).
(c) Raman spectra of liquid (upper blue) and solidified (lower red) DPHA.
The spectra are normalized by the peak intensity of $\delta$(C--H$_2$) of $\alpha$--C used as an internal reference.
}
\end{center}
\end{figure*}

As a material for proof-of-concept, we used acrylate oligomer, di-pentaerythritol hexaacrylate (DPHA) (Fig.~\ref{f2}(a), inset), which is a popular cross-linker material \cite{Kawata:2001hw,Maruo:1997ct}.%
DPHA is viscous liquid in room temperature and has six conjugated acryloyl moieties which become radical sites upon the cleavage of the C=C and C=O double bonds.
The absorption spectrum of DPHA is shown in Fig.~\ref{f2}(a).
A pronounced absorption is seen in the DUV wavelength region from 200 to 230\,nm, attributed to C=C and C=O functional groups in DPHA, while no absorption is seen at a wavelength of 400\,nm where our femtosecond laser for TPP-excitation is tuned.

We focused 400\,nm-femotosecond laser into the liquid oligomer and scan the focus spot along with the surface of a deer model (Fig.~\ref{f2}(b), left panel).
After the scan is completed, the sample was taken out of the residual liquid oligomer and observed by SEM.
The result is shown in the right panel in Fig.~\ref{f2}(b).
The deer sculpture stands with the shape well reproducing the CAD model (left panel).
The length of body is about 10\,$\mu$m.
The horns and the branches are free-standing with a diameter of about 300\,nm, demonstrating the capability to build high-aspect-ratio structures.

To investigate how DPHA undergoes chemical structural changes upon the laser irradiation, we measured Raman spectra of DPHA before and after the laser irradiation (details in the Methods).
The measured Raman spectra are shown in Fig.~\ref{f2}(c) with the lines in blue and red indicate the spectra before and after laser irradiation, respectively.
The Raman spectrum of DPHA shows four distinctive peaks (blue spectrum).
The peaks at 1402\,cm$^{-1}$ is assigned to the in-plane deformation mode of methylene group $\delta$(CH${_2}$); 1460\,cm$^{-1}$, the deformation mode of $\delta$(C--H${_2}$) in $\alpha$--C; 1630\,cm$^{-1}$ and 1720\,cm$^{-1}$, the stretching modes of C=C and C=O, respectively \cite{Edwards:2006gr,Willis:1969il}.
Assuming C--H${_2}$ in $\alpha$--C is not involved in polymerization reaction due to the transparency (no absorption) against DUV light, we used the peak intensity of $\delta$(C--H${_2}$) in $\alpha$--C at 1460\,cm$^{-1}$ as an internal reference to normalize Raman intensities.
As seen from Fig.~\ref{f2}(c), we observed a drastic decrease in the peak intensity of C=C at 1630\,cm$^{-1}$, indicating the consumption of C=C bond to form C--C networks.
The C=C consumption is accompanied by cleavage of C=O bond that is conjugated with the C=C bond \cite{Baldacchini:2009ib}, which manifests itself as a simultaneous decreasing of C=O peak at 1720\,cm$^{-1}$.
The decreasing intensity of CH${_2}$ band at 1402\,cm$^{-1}$ is explained by the cleavage of C=C bond connecting the CH${_2}$ group.

The applied laser intensity to obtain the deer sculpture was 300\,kW/cm$^2$ with the exposure time of 4\,ms at each scanning step.
The laser intensity required for initiating TPP process was found to be seven times smaller than the value used for previous NIR-excited-TPP for the same DPHA but added with photo-initiator and a photo-sensitizer (2.1\,MW/cm$^2$) \cite{Ushiba:2014id}.
In case of NIR excitation, the photo-initiator occupies only 1\,wt\% of monomer/cross-liner solution \cite{Ushiba:2014id}, whereas in case of DUV-TPP, all oligomer molecules inside the laser focus work as initiator.
The highly concentrated radicals generated by DUV absorption explains the unexpectedly high efficiency of DUV-TPP.

Previously, Parkatzidis \textit{et al.} reported TPP of pre-synthesized gelatin methacrylamide without addition of initiators using an excitation wavelength of 520\,nm \cite{Parkatzidis:2018ca}.
In the previous work using 520\,nm, the reported applied laser intensity at the focus was on the order of tens of TW/cm$^2$ \cite{Parkatzidis:2018ca}, which is more than eight orders of magnitudes higher than that we applied for polymerizing methacrylate oligomers with 400\,nm excitation.
In the former case, the two-photon energy of 520\,nm laser (i.e. 260\,nm) is hardly overlaps with the molecular absorption \cite{Parkatzidis:2018ca}, while in our case, two-photon energy of 400\,nm (i.e. 200\,nm) falls at the peak position of absorption (Fig.~\ref{f2}(a)).
The huge difference in required laser intensity indicates that the efficiency of two-photon polymerization is highly enhanced by using two-photon excitation energy consistent with one-photon absorption energy of molecules.



\begin{figure*}
\begin{center}
\includegraphics*[]{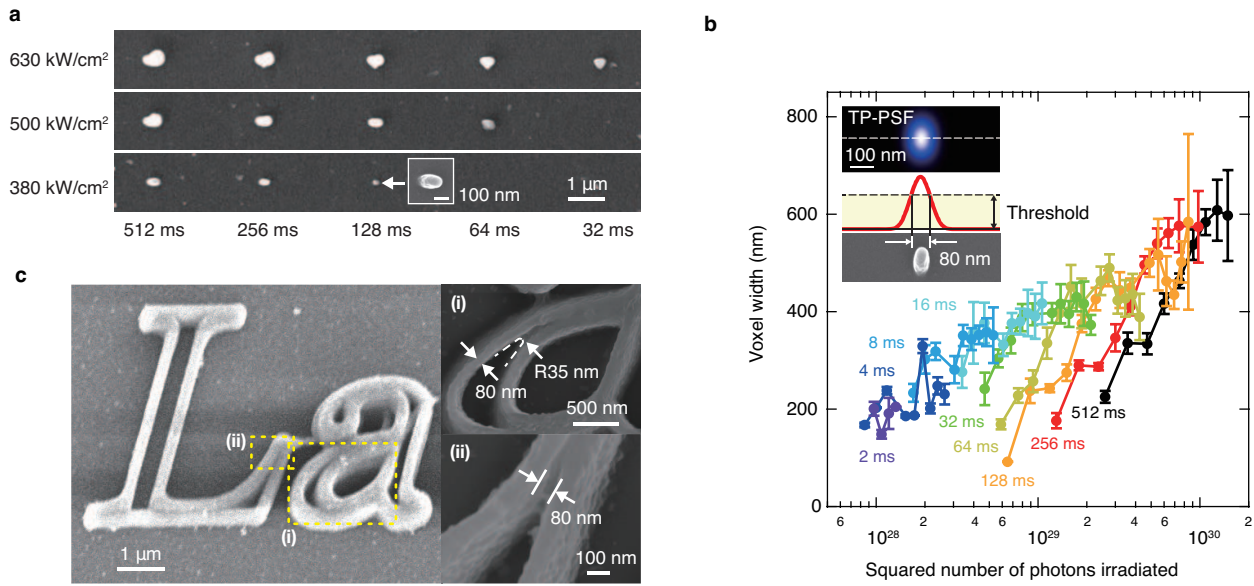}
\caption{\label{f3}
Spatial resolution.
(a) Scanning electron micrograph of voxels generated at different exposure times and laser power.
(b) Voxel width vs the squared number of photons irradiated.
The horizontal coordinate is logarithmic to show the non-linear behavior of TPP.
The inset shows two-photon point spread function (TP-PSF) image and line profile calculated numerically in condition with a wavelength of 400\,nm and objective lens NA of 1.4. The SEM image of the smallest voxel are also shown.
The illustrated threshold height is not the actual but a guide to eye. 
(b) SEM image of ``La'' letters. Regions marked by (i) and (ii) are magnified and shown in the right top and bottom panels, respectively.
}
\end{center}
\end{figure*}

To rationalize laser intensity and resolution in DUV-TPP, we systematically measure the voxel size at different irradiation intensities and times.
The SEM images of voxels are shown in Fig.~\ref{f3}(a) for the exposure power of 380, 500 and 630\,kW/cm$^2$ and time of 512, 256, 128, 64, and 32\,ms.%
The polarization of the excitation laser was linear and horizontally oriented, resulting in slightly elongated shape of voxels along with the polarized direction.
It is clearly seen that the size of voxel is smaller for lower laser intensity and shorter exposure time.
The smallest voxel was 80\,nm in width, which was obtained for the laser intensity of 380\,kW/cm$^2$ and the exposure time of 128\,ms.
Below 380\,kW/cm$^2$, we were unable to find a voxel on the substrate, partly because smaller voxels would have been rinsed off during the cleaning process before SEM observation.
For laser intensity of 630\,kW/cm$^2$ and the exposure time of 512\,ms, the voxel size increased as much as 600\,nm, which is larger than the size of laser spot given by the wavelength of 400\,nm and NA1.4.
Such large voxel was produced plausibly as a result of diffusion of radicals out of the laser focus \cite{Gleeson:2009jo, Zhou:2015ju}.

In Fig.~\ref{f3}(b), we plotted the voxel widths as a function of the squared number of photons irradiated to sample for different exposure time and intensities.
The voxel width phenomenologically lies linear-proportional to the logarithm of the squared number of photons, indicating a strong non-linearity.
For the fixed exposure time, the voxel width increases with increasing numbers of photon dose.
We numerically calculated two-photon point spread function (TP-PSF) with the condition of wavelength of 400\,nm and objective NA of 1.4.
The resulting image of TP-PSF and the sectional line profile at the center are shown in the inset in  Fig.~\ref{f3}(b).
The TP-PSF has a full-width at half-maximum of 139\,nm and 99\,nm in parallel and perpendicular to the polarization direction, respectively.
The observed smallest voxel width of 80\,nm falls to a fraction of the calculated TP-PSF width of 99\,nm.
Considering the thresholding effect in photo-polymerization \cite{Kawata:2001hw,Zhou:2015ju} (also illustrated in the inset), the experimentally obtained voxel size of 80\,nm reasonably agreed with the value numerically predicted from PSF.


To evaluate the spatial resolution of DUV-TPP, we show in Fig.~\ref{f3}(c) the SEM images of ``La'' letters drawn by DUV-TPP.
The edge of the fabricated structure is remarkably sharp.
Regions marked by (i) and (ii) are magnified and shown in the right upper and lower panels, respectively.
The line width of 80\,nm and a curved feature with a radius of curvature of 35\,nm were observed in panel (i).
In panel (ii), the space resolved between two crossing lines was about 80\,nm, reasonably consistent with the smallest voxel size observed in Fig.~\ref{f3}(a).
From these results, we concluded the finest resolved space is about 80\,nm wide, corresponding to $\lambda/5$ of excitation wavelength.
The value of spatial resolution achieved with 400\,nm excitation is superior to those previously reported for NIR-excitation \cite{Kawata:2001hw} and 520\,nm excitation \cite{Malinauskas:2010bm}.


Next, we apply initiator-free DUV-TPP to another class of material, inorganic metal oxides.
Zirconium dioxide (ZrO$_2$) exhibits large dielectric constant, known as high-$\kappa$ material, and has potentially important applications in optical \cite{Liang:2009cr} and electronic devices \cite{Park:2011iq}.
Titanium dioxide (TiO$_2$) also has a wide range of applications including photocatalysis \cite{Fujishima:2008co}.
There are numbers of techniques to make a planer thin film of metal oxides \cite{Liang:2009cr}, but a technique to form 3D nanostructures will be demanded, for example, to integrate metal oxides nanostructures in electronic devices.

\begin{figure*}
\begin{center}
\includegraphics*[]{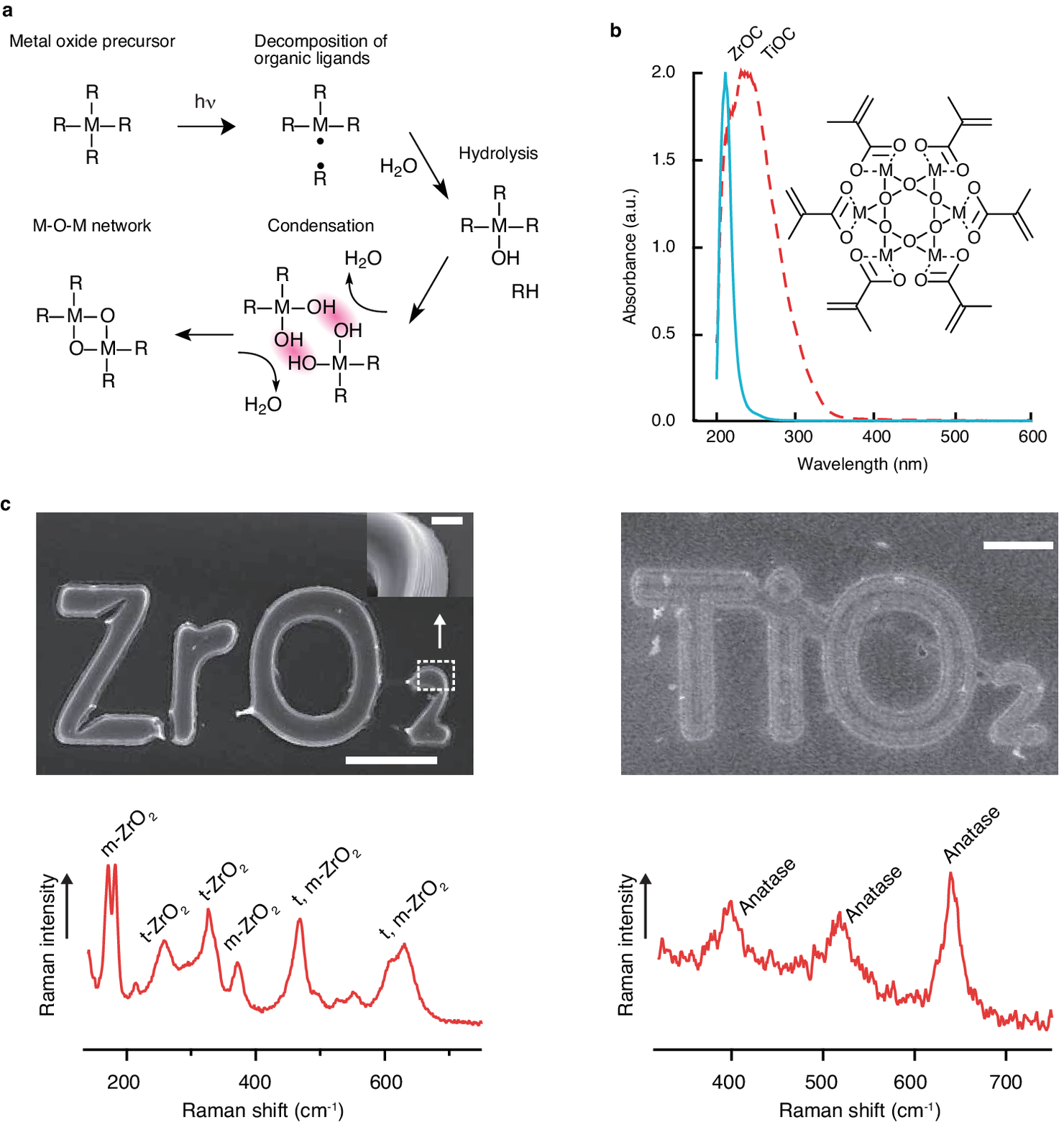}
\caption{\label{f4}
Metal oxide microstructures fabricated by TPP. 
a, Diagram showing reaction pathway from metal oxide precursor to metal oxide network.
b, Absorption spectra of metal-oxio-clusters (MOCs).
The spectra for ZrOC and TiOC are shown in blue slid line and red dashed line, respectively.
c, SEM images of microstructures made by ZrO$_2$ (upper left), TiO$_{2}$ (upper right), and corresponding Raman spectra taken from a part of structures (lower panels).
The scale bar, 5\,$\mu$m (inset 500\,nm).
}
\end{center}
\end{figure*}

Figure~\ref{f4}(a) explains the reaction pathway to synthesize metal oxides.
It starts from metal oxo cluster (MOC) as a precursor (Method).
The MOC in solution phase has organic ligands surrounding the metal cation.
Photo excitation of ligands promotes decomposition of organic ligands.
The generated radicals leads hydrolysis and condensation, and metal-oxygen-metal (MOM) network is created.
We patterned the MOM structures by DUV-TPP.
No photo-initiator and sensitizers were added to MOC solution.
The solidified MOM structures were taken out of residual unsolidified MOC solution and finally thermally annealed to crystalize metal oxides.

The absorption spectrum of synthesized MOC solution is shown in Fig.~\ref{f4}(b) for ZrOC and TiOC.
ZrOC has an absorption peak at 210\,nm, attributed to C=C bond from ligand MMA.
TiOC has an absorption originated from the C=C bond from ligand MMA, but the peak wavelength is slightly shifted to 230\,nm and the width is broader.
In both cases, the absorption band of the MOC overlaps with the two-photon excitation energy with a laser wavelength centered at 400\,nm. 

Figure~\ref{f4}(c) shows the SEM image of the fabricated structures with letters ``ZrO$_2$'' and ``TiO$_2$'' from the respective ZrOC and TiOC solutions.
The inset of ``ZrO$_2$'' image shows the magnified portion in the side wall of the letter ``2''.
A layered feature in vertical direction is seen on the wall of the letter, demonstrating the 3D capability.
The structures were slightly deformed during thermal annealing.
Further optimization is needed to avoid deformation and cracks during annealing process \cite{Kozuka:2006eq}.
To identify the chemical components of the obtained structures, we took Raman spectra from a part of the structures.
The result is shown in the bottom panels in Fig.~\ref{f4}(c).
The spectrum in the left panel shows characteristic Raman peaks.
Comparing these peaks to the literatures \cite{Li:2001fh}, we found the peaks at 269 and 310\,cm$^{-1}$ are assigned to the Raman modes of tetragonal phase t-ZrO$_2$, Eg, and B1g, respectively \cite{Li:2001fh}.
The other peaks at 178, 189, and 380\,cm$^{-1}$ are assigned to the monoclinic phase m-ZrO$_2$ \cite{Li:2001fh}.
Consequently, the fabricated ``ZrO$_2$'' structure is composed of polycrystalline ZrO$_2$ that has both tetragonal and monoclinic phases.
The remaining peaks at 469 and 640\,cm$^{-1}$ are reported for both t- and m-ZrO$_2$ \cite{Li:2001fh}.
In the measured spectrum, the Raman intensity at 640\,cm$^{-1}$ is stronger than that at 469\,cm$^{-1}$, which is an indication that the obtained structure is t-ZrO$_2$-phase rich.
The dominant phase is affected by the annealing temperature of the crystallization process \cite{Li:2001fh}.
No remaining peak of ligand MMA was observed in the spectrum.
The Raman spectrum from ``TiO$_2$'' structure shows three distinct peaks, 4008, 517, and 639\,cm$^{-1}$.
These peaks are all consistent with the reported Raman modes in TiO$_2$ anatase crystal \cite{Berger:1993eu}.
The results of Raman spectroscopy show the obtained structures are indeed ZrO$_2$ and TiO$_2$ crystals.
To the best of our knowledge, this is the first to build structures made of inorganic metal compound by direct laser writing.


In order to gain further insight about applicability and limitation of DUV-TPP, we have tested polymerization of MMA monomers with the experimental conditions similar to the one used for MMA oligomers.
As a result, however, we were unable to find any structure remaining on the substrate after laser irradiations.
Unlike acrylate oligomer having six acryloyl moieties within a molecule, MMA monomer has only a single set of C=C and C=O.
Since there are a limited number of double bonds available to form crosslinks with each other, MMA monomer has less chance to form polymer matrix during a given lifetime of radicals than MMA oligomer does.
Applying DUV-TPP to simpler molecules having a small numbers of radical sites would become possible by increasing monomer concentrations or reducing a solution viscosity, which will be a future work.

Another limitation of DUV-TPP from material point of view is the molecules with conjugated system.
It is known that the electronic transition energy of conjugated system becomes lower when the number of conjugated double bond increases.
When polymerized molecules absorbs visible light, one-photon polymerization dominates over two-photon polymerization, which spoils 3D fabrication capability.
As a solution, single-pulse polymerization \cite{Mills:2013ija} may work.

There are reports of cytotoxicity during cell culturing on gelatin hydrogel scaffolds fabricated with photo-initiators \cite{Li:2013bd,Williams:2005ds}.
While chosing appropriate initiator is a crucial issue in NIR-excited TPP, initiator-free DUV-TPP is a promising alternative approach for applications that require high degree of bio-compatibility and nontoxicity, such as cell culturing scaffolds, regenerative medicines, drag delivery containers \cite{Xing:2015ey,Lee:2019ki}.
Experiments are ongoing to fabricate structures with collagen type I molecules using DUV-TPP, and obtained solidified structures after irradiation of 400\,nm laser (Supplementary Information and Fig.~S1).

As demonstrated with metal oxides structures, DUV-TPP widens the choice of materials for 3D nanofabrications.
Using synthesized ligand MMA as cross-linker, DUV-TPP is potentially applicable to 3D nanofabrications of wide variety of inorganic materials including metals \cite{Tanaka:2006hp} and SiO$_2$ \cite{Kotz:2017ez}.
With the material versatility, an interesting extension is fabrication of multi-material nanodevices \cite{Lind:2017gn,Kwon:2018kq,Dietrich:2018do}


Finally, multi-photon excitation with visible laser offers facile and convenient way to realize maskless 3D printing with equivalent operating wavelength in DUV.
This enables rapid prototyping or testing of arbitrary macro- and microstructures and devices for many applications in both industry and fundamental researches.
Since laser intensity required is relatively small in DUV-TPP, writing throughput can be improved by parallelization using multi-spots scheme.
Strong nonlinearity involved in polymerization process would be utilized to further improve the spatial resolution to the level comparable to the state-of-the-art DUV lithography and EBL \cite{Gan:2013ed,Xing:2007gu}.
In summary, we have described DUV photo-polymerization with two-photon excitation at visible wavelength.
Without using photo-initiators and sensitizers, direct multi-photon excitation of molecular native absorption has shown to induce photo-polymerizations.
This expands the scope of TPP-polymerizable materials from traditional organic materials to inorganic materials as demonstrated by fabrication of both acrylate and metal oxides nanostrucutures in 3D.

\section{acknowledgments}
This work was supported partly by the JSPS KAKENHI Grant Numbers JP17K05076, and JSPS Core-to-Core Program on Advanced Nanophotonics.

\section{Author contributions}
A.T. and K.F. conceived the project and designed the experiments.
A.N. performed the experiments and R.O. provided the calculation data.
S.K. participated in planning the experiments and provided conceptual input.
The manuscript was written by A.T., A.N. and K.F.


\section{Methods}
\paragraph{Optical configuration of two-photon polymerizations}
An experimental setup is shown in Fig.~\ref{f1}(d).
A Ti:sapphire oscillator system (Tsunami, Spectra Physics) was used as a light source.
The fundamental 800\,nm NIR laser beam (80\,fs pulse width, 82\,MHz repetition rate, 10\,nm full-width at half- maximum (FWHM) bandwidth) was focused to a $\beta$-BaB$_{2}$O$_{4}$ (BBO) crystal using a lens (f=50\,mm) to provide a visible laser pulse having a center wavelength at 400\,nm.
The frequency-doubled laser beam was expanded and collimated by a lens (f=100\,mm).
The laser was passed through a short-pass filter to cut the fundamental 800\,nm beam, and the visible laser coupled into an Olympus IX71 inverted microscope on which a sample cell is set.
A 100$\times$ oil immersion objective (NA 1.4, Zeiss) was used to focus the visible light into sample solution.
The sample stage moves three-dimensionally in $XYZ$ using a piezo stage (P-517, Physik Instrument) which was controlled by home-written programs.

\paragraph{Experimental procedures for two-photon polymerizations}
\subparagraph{Acrylate sample preparation}
For the two-photon polymerizations of acrylate oligomer shown in Fig.~\ref{f2}, we used di-pentaerythritol hexaacrylate (DPHA) (Kyoeisha Chemical Co., Ltd.), which is a popular acrylate oligomer.
The stock solution of DPHA was used as received.
The absorption spectra was measured using UV-Vis-NIR spectrometer (UV-3600, Shimadzu).
A droplet of DPHA solution was sealed between a pair of two coverslips (Matsunami glass) with a 50\,${\mu}$m-thick silicone sheet sandwiched as a spacer.

\subparagraph{Metal oxio-cluster (MOC) preparation}
Zr oxio-cluster (ZrOC) was prepared following the procedures described in the literature\cite{Stehlin:2014be}.
Briefly, Zr alkoxide (zirconium(IV) propoxide (ca.~70\% in 1-Propanol, Tokyo Chemical Industry Co., Ltd.) precursor was mixed with methyl methacrylate (MMA) (Tokyo Chemical Industry Co., Ltd.) at the molar ratio of Zr:MMA=1:8. After stirring of five minutes, a volume of 20\,$\mu$L of 1-propanol (Tokyo Chemical Industry Co., Ltd.) was added.
After ten minutes of stirring, de-ionized water was added at a molar ratio of Zr:H2O=1:20.
Finally, an additional volume of 20\,$\mu$L 1-propanol was added to adjust the viscosity of solution.

Ti oxio-cluster (TiOC) was prepared with the same procedure except for using Ti alkoxide (titanium (IV) isopropoxide, 97\%, Aldrich) in place of Zr alkoxide.

\subparagraph{Fabrications}
To produce a three dimensional structure, we first generate a computer model using a computer-aided design (CAD) software.
The CAD model is then converted to a batch of G-codes using a slicer software (Cura), and fed into a scan controller to move the sample stage.
After the scan is completed, residual liquid oligomer is rinsed off using ethanol followed by a supercritical cleaning that uses CO$_{2}$ supercritical fluid having extremely small viscosity to reduce a risk of collapsing nanostructures during cleaning process \cite{Maruo:2009bb}.

For fabrication of metal oxides structures, the experimental conditions used for DUV-TPP were the same as the one used for the fabrication of acrylic polymer except for the laser intensity 130\,kW/cm$^2$ and exposure time of 4\.ms per voxel.
As a solvent for rinsing the residual liquid MOC, cyclohexanone (Tokyo Chemical Industry) was used.
After cleaning, the MOM structure was thermally annealed at a temperature of 600C$^{\circ}$ for two hours to completely decompose the organic ligands and to crystalize the metal oxide structure.

\subparagraph{SEM observation}
To observe the fabricated structures, the sample was coated with a thin platinum layer of 2\,nm thick.
A field emission scanning electron microscope (S-4800, Hitachi) was used.

\subparagraph{Raman microscopy characterization}
Raman spectra were taken using a Raman microscope (Raman-11, Nanophoton).
The excitation laser wavelength was 532\,nm and a 100$\times$ objective lens (NA0.9, Nikon) was used.
The laser power was 100\,mW and the exposure time was 1\,s.




\end{document}


\preprint{Preprint submitted to Nature Photonics}

\title{Supplementary Information for \\``Multiphoton-excited DUV photolithography for 3D nanofabrication''}



\author{Atsushi Taguchi}
\email[]{taguchi@es.hokudai.ac.jp}
\affiliation{Department of Applied Physics, Osaka University, Suita, Osaka 565-0871, Japan}
\altaffiliation{Present address: Research Institute for Electronic Science, Hokkaido University, Sapporo, Hokkaido 001-0020, Japan}
\author{Atsushi Nakayama}
\affiliation{Department of Applied Physics, Osaka University, Suita, Osaka 565-0871, Japan}
\affiliation{Advanced Photonics and Biosensing Open Innovation Laboratory, AIST-Osaka University, Suita, Osaka 565-0871, Japan}
\author{Ryosuke Oketani}
\affiliation{Department of Applied Physics, Osaka University, Suita, Osaka 565-0871, Japan}
\author{Satoshi Kawata}
\affiliation{Department of Applied Physics, Osaka University, Suita, Osaka 565-0871, Japan}
\affiliation{Serendip Research, Suita, Osaka 565-0871, Japan}
\author{Katsumasa Fujita}
\email[]{fujita@ap.eng.osaka-u.ac.jp}
\affiliation{Department of Applied Physics, Osaka University, Suita, Osaka 565-0871, Japan}
\affiliation{Advanced Photonics and Biosensing Open Innovation Laboratory, AIST-Osaka University, Suita, Osaka 565-0871, Japan}
\affiliation{Transdimensional Life Imaging Division, Institute for Open and Transdisciplinary Research Initiatives, Osaka University, Suita, Osaka 565-0871, Japan}





\maketitle 


\section{Absorption of representative functional groups}
Table \ref{fT1} shows the absorption characteristics of representative functional groups \cite{Calvert:1966tt,Gillam:1970vt}.

\begin{table}[htb]
\caption{Absorption of representative functional groups}\label{fT1}
\begin{center}
\begin{tabular}{lcc}
\hline
Functional group & \phantom{10\em}$\lambda_\mathrm{max}$ (nm)\phantom{10\em} & $\epsilon_\mathrm{max}$ (dm$^3$ mol$^{-1}$ cm$^{-1}$) \\
\hline\hline
C=C & 162 & 10,000 \\
 & 170 & 16,500 \\
 & 174 & 16,500 \\
 & 183 & 250\\
\hline
C=O & 180 & 10,000 \\
 & 295 & 10 \\
\hline
C=S & 240 & 10,000 \\
 & 490 & 10 \\
\hline
-NO$_2$ & 185 & Strong \\
 & 210 & 10,000 \\
 & 278 & 10 \\
\hline
-N=N- & $\sim$260 & Strong \\
 & 347 & 15 \\
\hline
Benzen ring & 180 & 100,000 \\
 & 200 & 6,300 \\
 & 255 & 200 \\
\hline
C=C--C=C & 217 & 21,000 \\
\hline
C=C--C=C--C=C & 258 & 35,000 \\
\hline
\end{tabular}
\end{center}
\label{default}
\end{table}%

\section{Collagen}

While biocompatible photo-polymerizing polymers have been extensively studied for culturing cells using 3D scaffolds, cytotoxicity of photo-initiator has been reported \cite{Li:2013bd,Ovsianikov:2011ku,Williams:2005ds,Ovsianikov:2014dv}.
To reduce cytotoxicity, photoinitiator-free polymerization has been reported for poly(ethylene glycole) diacrylate with one-photon excitation using 248\,nm excimer laser \cite{Farkas:2017iu} and for gelatin methacrylamide (GelMA) biopolymer using 520\,nm femtosecond pulsed laser excitation \cite{Parkatzidis:2018ca}.
In both cases, initiator-free polymerizations were realized by pre-modifying monomers to bear acrylate groups which work as a cross-linking site upon photo-excitation.
In our case, unlike the previous reports, we used pure collagen molecule as monomer without using pre-modification nor chemical synthetic process.

\begin{figure*}
\begin{center}
\includegraphics*[]{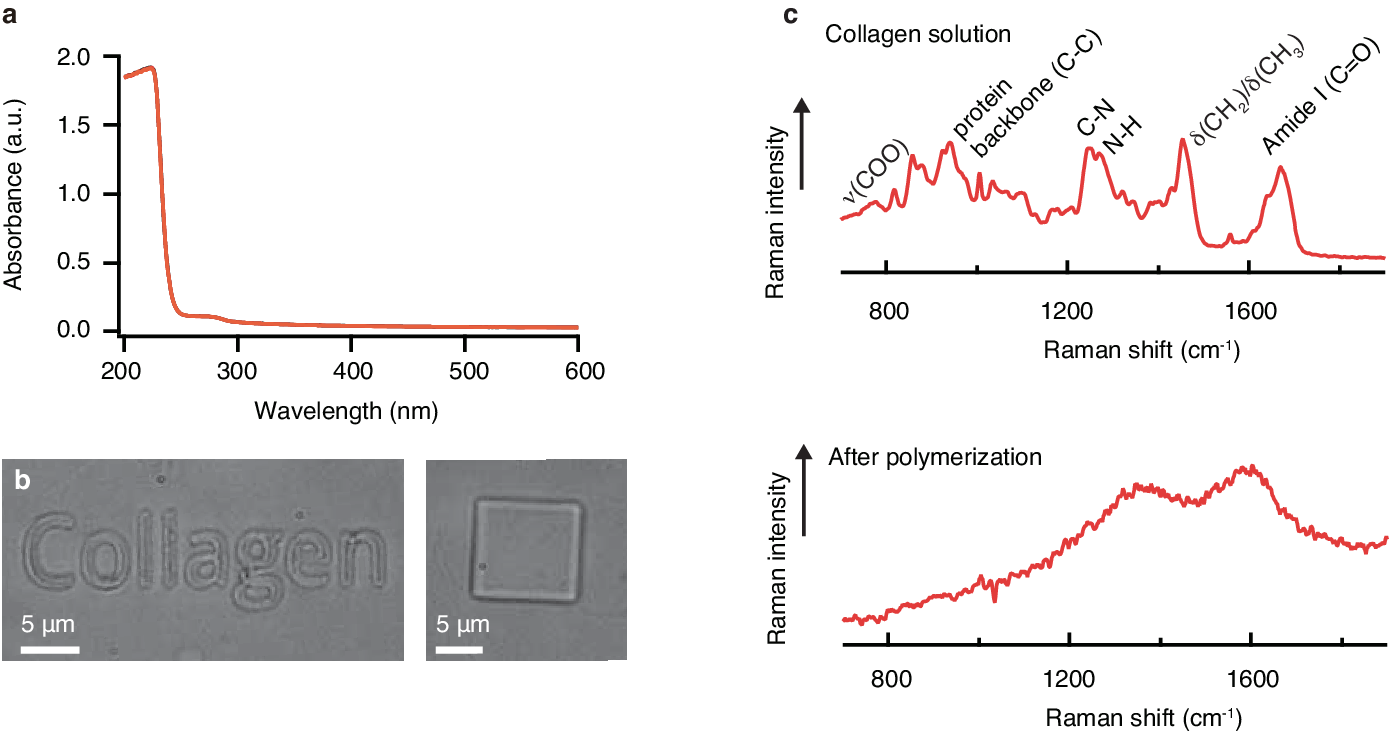}
\caption{\label{fS1}
Microstructures made of collagen protein.
(a) Optical microscope images of collagen polymerized by DUV-TPP: letters ``Collagen'' (left) and a rectangular structure (right).
(b) Raman spectra of collagen before polymerization (top) and that from the solidified rectangular structure after polymerization (bottom).
}
\end{center}
\end{figure*}

Collagen type I from rat tail was purchased from Sigma and used without further purification.
Collagen solution with a concentration of 3\,mg/mL was prepared by dilution in 0.2\,M acetic acid solution \cite{Basu:2005bc}.
The absorption spectrum of pre-polymerized collagen solution is shown in Fig.~\ref{fS1}(a).%
A pronounced absorption band is seen at the wavelength between 200 and 220\,nm, while no absorption is recognized in the visible wavelength including 400\,nm at which our excitation laser is tuned.
Collagen protein is composed of amino acids, one third of which is glycine and 15--30\% of which are proline and 4-hydroxyproline.
The strong absorption of collagen at around 200\,nm is associated with proline \cite{GRATZER:1963th}.

The experimental procedure of TPP is the same as described in the main text.
The excitation laser intensity and exposure time was set at 1.7\,MW/cm$^2$ and 4\,ms per each scanning step, respectively.
The applied laser intensity was eight orders of magnitude smaller than that used for polymerizing GelMA using 520\,nm excitation \cite{Parkatzidis:2018ca}.

Figure \ref{fS1}(b) shows the optical microscope images of structures obtained after 400\,nm-femtosecond laser focus was scanned along with the surface of the pre-designed structures, one is letters ``Collagen'' and the another is a square box.
As seen in the figure, the collagen irradiated with a 400-nm pulse laser was solidified.
The designed thickness of the letter structure is 4.2\,$\mu$m, and the box structure is 5\,${\mu}$m thick and 10\,${\mu}$m square.


Raman spectra of collagen before and after polymerization are shown in Fig.~\ref{fS1}(c) in the top and bottom panels, respectively.
For the Raman spectrum before polymerization, the shape and spectral position is well consistent with that reported in literature \cite{Janko:2010fi}: C-C protein backbone at 939\,cm$^{-1}$; $\nu$(CCO) at 767\,cm$^{-1}$; C-N at 1246\,cm$^{-1}$; $\delta$(N-H) the deformation band of the amide III at 1271\,cm$^{-1}$; and $ \nu$(CO) the C=O stretching vibration of amide I at 1668\,cm$^{-1}$. The 1451\,cm$^{-1}$ band is assigned to methyl $\delta$(CH$_3$) and methylene $\delta$(CH$_2$) deformation vibrations, which is also found in the literature \cite{Janko:2010fi}.
After polymerization, the characteristic Raman peaks of collagen disappeared and two broad peaks remain in the Raman spectrum (the lower panel).
Judging from the Raman spectral shape with two broad peaks, the obtained Raman spectrum after polymerization is most likely from amorphous carbon \cite{Ferrari:2001db}.
Further experiments are needed with careful optimization to obtain collagen that is polymerized but not fully carbonized, which is currently ongoing.

\begin{acknowledgments}

\end{acknowledgments}
